\documentclass{elsart}

\usepackage{graphicx} 
\usepackage{epsfig}   
\usepackage{dcolumn}  
\usepackage{color}

\graphicspath{{ps}}

\newcommand{\GeV}{\mbox{GeV}}
\newcommand{\MeV}{\mbox{MeV}}
\newcommand{\fb}{\mbox{fb}}

\begin{document}

\begin{frontmatter}

\epsfysize2.cm
\hspace{-9.5cm}
\begin{flushright}
\vskip -1.5cm
\noindent
\hspace*{9.0cm}Belle Prerpint 2006-36\\
\hspace*{9.0cm}KEK   Preprint 2006-53\\
\hspace*{9.0cm}hep-ex/0611045
\end{flushright}

\title{ \boldmath \quad\\[0.5cm] A Search for the Rare Leptonic Decays $B^+\to \mu^+\,\nu_\mu$ and $B^+\to e^+\,\nu_e$}

\collab{Belle Collaboration}
  \author[Shinshu]{N.~Satoyama}, 
  \author[KEK]{K.~Abe}, 
  \author[KEK]{I.~Adachi}, 
  \author[Tokyo]{H.~Aihara}, 
  \author[BINP]{D.~Anipko}, 
  \author[Sydney]{A.~M.~Bakich}, 
  \author[Melbourne]{E.~Barberio}, 
  \author[BINP]{I.~Bedny}, 
  \author[Protvino]{K.~Belous}, 
  \author[JSI]{U.~Bitenc}, 
% ...

\address[BINP]{Budker Institute of Nuclear Physics, Novosibirsk, Russia}
% ...
\address[KEK]{High Energy Accelerator Research Organization (KEK), Tsukuba, Japan}
% ...
\address[Protvino]{Institute for High Energy Physics, Protvino, Russia}
% ...
\address[JSI]{J. Stefan Institute, Ljubljana, Slovenia}
% ...
\address[Melbourne]{University of Melbourne, Victoria, Australia}
% ...
\address[Shinshu]{Shinshu University, Nagano, Japan}
% ...
\address[Sydney]{University of Sydney, Sydney, NSW, Australia}
% ...
\address[Tokyo]{Department of Physics, University of Tokyo, Tokyo, Japan}
% ...

\vspace{1.0cm}

\begin{abstract}
We present a search for the decays $B^+\to \mu^+\nu_\mu$ and $B^+\to e^+\nu_e$
in a $253\,\textrm{fb}^{-1}$
data sample collected at the $\Upsilon(4S)$ resonance with the Belle detector 
at the KEKB asymmetric-energy $B$ factory. 
We find no significant evidence for a signal and set 90\% confidence level
upper limits of 
${\mathcal B}(B^+\to\mu^+\nu_\mu) < 1.7 \times 10^{-6}$ and 
${\mathcal B}(B^+\to e^+\nu_e) < 9.8 \times 10^{-7}$.

\end{abstract}

\begin{keyword}
Leptonic, B decay
\PACS 13.25.Hw, 14.20.Lq, 14.40.Nd
\end{keyword}
\end{frontmatter}



{\renewcommand{\thefootnote}{\fnsymbol{footnote}}}
\setcounter{footnote}{0}

\section{Introduction}
\label{sec:Introduction}

The purely leptonic decay $B^{+}\to\ell^{+}\nu_{\ell}$ 
(charge conjugate states are implied throughout the paper)
is of particular interest
since it provides a direct measurement of the product of 
the magnitude of the Cabibbo-Kobayashi-Maskawa matrix
element, $|V_{ub}|$~\cite{Cabibbo:1963yz},
and the $B$ meson decay constant, $f_{B}$.
In the Standard Model (SM) the branching fraction of the decay 
$B^{+}\to\ell^{+}\nu_{\ell}$  is given as
\begin{equation}
 \label{eq:BR_B_taunu}
{\mathcal B}(B^{+}\to\ell^{+}\nu_{\ell}) = \frac{G_{F}^{2}m_{B}m_{\ell}^{2}}{8\pi}\left(1-\frac{m_{\ell}^{2}}{m_{B}^{2}}\right)^{2}f_{B}^{2}|V_{ub}|^{2}\tau_{B},
\end{equation}
where $G_{F}$ is the Fermi coupling constant, $m_{\ell}$ and $m_{B}$ are
the charged lepton and $B$ meson masses, and $\tau_{B}$ is the $B^{+}$ lifetime.
The expected branching fractions are  $(4.7\pm 0.7) \times 10^{-7}$
for $B^+\to\mu^+\nu_\mu$ and $(1.1 \pm 0.2) \times 10^{-11}$ for $B^+\to e^+\nu_e$
assuming $|V_{ub}| = (4.39 \pm 0.33) \times 10^{-3}$ determined from inclusive
charmless semileptonic $B$ decay data~\cite{HFAG}, $\tau_{B} = 1.643\pm 0.010$ ps~\cite{HFAG},
and $f_B = 0.216\pm 0.022\,\GeV$
obtained from lattice QCD calculations~\cite{Gray:2005ad}.
However, non-SM physics could yield larger branching
fractions~\cite{higgs}. 

The Belle experiment has recently found the first evidence for the purely leptonic decay
$B^{+}\to\tau^{+}\nu_{\tau}$~\cite{Ikado:2006un}.
The other purely leptonic $B$ decays, $B^+\to\mu^+\nu_\mu$ and $B^+\to e^+\nu_e$,
have not yet been observed. The most stringent current
upper limits for these modes are
${\mathcal B}(B^+\to\mu^+\nu_\mu) < 6.6 \times 10^{-6}$~\cite{babar-2004} and
${\mathcal B}(B^+\to e^+\nu_e) < 1.5 \times 10^{-5}$~\cite{CLEO-1995}.
Preliminary limits of
${\mathcal B}(B^+\to\mu^+\nu_\mu) < 2.0\times 10^{-6}$~\cite{Belle-ICHEP04} and 
${\mathcal B}(B^+\to e^+\nu_e) < 7.9\times 10^{-6}$~\cite{BaBar-ICHEP06}
are also available from the Belle and BaBar
collaborations, respectively.
In this paper, we present a search for the decays
$B^+\to\mu^+\nu_\mu$ and $B^+\to e^+\nu_e$.

\section{Data Set and Experiment}
\label{sec:data_exp}

The data used in this analysis were collected with the Belle detector
at the KEKB asymmetric-energy $e^{+}e^{-}$
collider~\cite{Kurokawa:2001nw}.
The sample has an integrated luminosity of $253\,\fb^{-1}$
accumulated at the $\Upsilon(4S)$ resonance,
at a center-of-mass (CM) energy of $10.58\,\GeV$
(on-resonance), and $28.1\,\fb^{-1}$
accumulated at a CM energy $60\,\MeV$ below 
the $\Upsilon(4S)$ resonance (off-resonance).

The Belle detector is a large solid-angle spectrometer based on
a $1.5\,$T superconducting solenoid magnet. Charged particle tracking and momentum
measurements are made with a three-layer,
double-sided silicon vertex detector (SVD) and a central drift
chamber (CDC). Identification of charged
hadrons is provided by a combination of three measurements: specific
ionization loss ($dE/dx$) in the CDC, photon yield in the aerogel threshold
Cherenkov counters (ACC), and time-of-flight information from a cylindrical
array of 128 scintillation counters (TOF). Photons are detected in an
electromagnetic calorimeter (ECL) system made of an array of 8736 CsI(Tl)
crystals surrounding the TOF system. The ECL is also used for electron
identification.
Muons are identified by a resistive plate
chamber system (KLM) located within the solenoid's external return yoke.

Particle identification for $e^{\pm}$ and $\mu^{\pm}$ is important
for this analysis.
Electron identification is based on the ratio of the cluster energy
in the ECL to the track momentum from the CDC and the SVD
$(E/p)$, the $dE/dx$ in
the CDC, the position and shower shape of the cluster in the ECL and
the response from the ACC. Muon identification is based on the hit
positions and the depth of penetration into the KLM.
The efficiency of electron
identification is over $90\,\%$ in the momentum range of this analysis
while the misidentification rate is below $0.5\,\%$. 
The muon identification efficiency is approximately $85\,\%$ in the momentum
range of this analysis with a misidentification rate of approximately $1\,\%$.
The Belle detector is described in detail
elsewhere~\cite{BelleDetector}.

A detailed Monte Carlo (MC) simulation, which fully describes the detector
geometry and response  based on GEANT~\cite{GEANT}, is
used to estimate the signal detection efficiency and to study the 
background. The MC samples are produced with the EvtGen event
generator~\cite{Lange:2001uf}.

\section{Signal Selection}
\label{sec:signal_selection}

We search for events in which there is one well-identified lepton.
The signal candidate muon or electron is required 
to pass tight particle identification requirements; lepton candidates
that can be associated to other tracks in the event to reconstruct 
$K^{0}_{\rm S}$ mesons or photon conversions are explicitly vetoed.
Signal lepton candidates are required to originate from the interaction
point (IP) and to have their polar angle, $\theta_\ell$, formed by the
lepton momentum with the detector axis
(opposite to the direction of the positron beam),
in the range $-0.5 < \cos\theta_\ell < 0.8$ for muons and
$-0.50 < \cos\theta_\ell < 0.85$ for electrons. 

As $B^{+}\to\ell^{+}\nu_{\ell}$ is a two-body decay,
the lepton has a fixed momentum in the signal $B$ meson ($B^{\rm sig}$)
rest frame, with $p_\ell^B$ equal to
approximately half of the $B$ meson mass, $p_\ell^B\sim m_B/2$.
The lepton momentum in the CM frame, $p_\ell^*$, is related to
$p_\ell^B$ by
\begin{equation}
  p_\ell^B  \simeq  p_\ell^*\left(1-\frac{|\vec p_{
   B^{\rm sig}}^{\,*}|}{m_B}\cos\theta_{\ell-B^{\rm sig}} \right)
  \label{eq:momentum}
\end{equation}
where $\vec p_{B^{\rm sig}}^{\,*}$ is the momentum of the $B^{\rm sig}$
in the CM frame and $\cos\theta_{\ell -B^{\rm sig}}$ 
represents the cosine of the angle between the directions of 
the signal lepton and $B^{\rm sig}$ in the CM frame.

Since the neutrino is not detected in the $B^{\rm sig}$ decay, we
obtain $\vec p_{B^{\rm sig}}^{\,*}$ by inclusive reconstruction of the companion
$B$ meson ($B^\mathrm{comp}$) recoiling against $B^{\rm sig}$. 
For the $B^\mathrm{comp}$ reconstruction, we use all detected photons and charged tracks, except
for the signal lepton candidate. 
$K^{0}_{\rm S}\to\pi^+\pi^-$ and $\gamma\to e^+e^-$ decays are fully
reconstructed, in order to correctly account for the momentum of tracks
originating from vertices displaced from the IP. 
The missing momentum in the laboratory frame that is
calculated by using all photons, charged tracks, and the signal lepton, is
assigned to the neutrino.
The quantity  $|\vec p_{B^{\rm sig}}^{*}|$ is approximately given by 
$\sqrt{E^2_{\rm beam} - m_B^2 }\simeq 0.32\,\GeV/c$ using the beam
energy in the CM frame ($E_{\rm beam}$),
while $\cos\theta_{\ell-B^{\rm sig}}$ is related to the angle between the
direction of the signal lepton and the momentum of
$B^{\rm comp}(\theta_{\ell-B^{\rm comp}})$ by 
$\cos\theta_{\ell-B^{\rm sig}} = - \cos\theta_{\ell-B^{\rm comp}}$. Thus,
equation~\ref{eq:momentum} can be expressed 
in terms of two measurable quantities, $p^{*}_\ell$ and
$\cos\theta_{\ell-B^{\rm comp}}$:
\begin{equation}
  p_\ell^B  \simeq  p_\ell^*(1+0.06\cos\theta_{\ell-B^{\rm
   comp}}).
   \label{eq:apploximation}
\end{equation}

Signal candidates are selected using the kinematic variables
$M_{\rm bc}  = \sqrt{E_\mathrm{beam}^2 - |{\vec p_{B}^{\,*}}|^2}$
and $\Delta E = E_B^* -E_\mathrm{beam}$, where 
$\vec p_{B}^{\,*}$ and $E_B^*$ are the momentum and energy
of $B^{\rm comp}$, all
variables being evaluated in the CM frame.
Events in the \emph{fit region}
that satisfy $5.10 \,\GeV/c^2 < M_{\rm bc} < 5.29\,\GeV/c^2$ and 
$-0.8~(-1.0)\,\GeV < \Delta E < 0.4\,\GeV$ for the muon (electron) mode
are kept for further analysis. 
A more restricted \emph{signal region} is also defined by the same
requirements on $\Delta E$ and by $M_{\rm bc} > 5.26\,\GeV/c^2$.

The dominant background arises from the continuum process 
$e^+e^-\to q\bar q\ (q=u,d,s,c)$
and semileptonic $B$ meson decays, mostly into charm 
final states ($B\to X_c\ell\nu$) with contributions from rare $b \to u$
modes ($B\to X_u \ell\nu$).
The missing momentum of continuum events is often due to undetected particles that
are outside the detector acceptance. In order to reduce such backgrounds, 
we require the transverse component 
of the missing momentum to be
greater than $1.75\,\GeV/c$, and the cosine of the polar angle 
to be less than $0.84$ ($0.82$) for the muon
(electron) mode, respectively.

\begin{figure}[tb]
  \begin{center}
   \includegraphics[scale=0.34,angle=0]{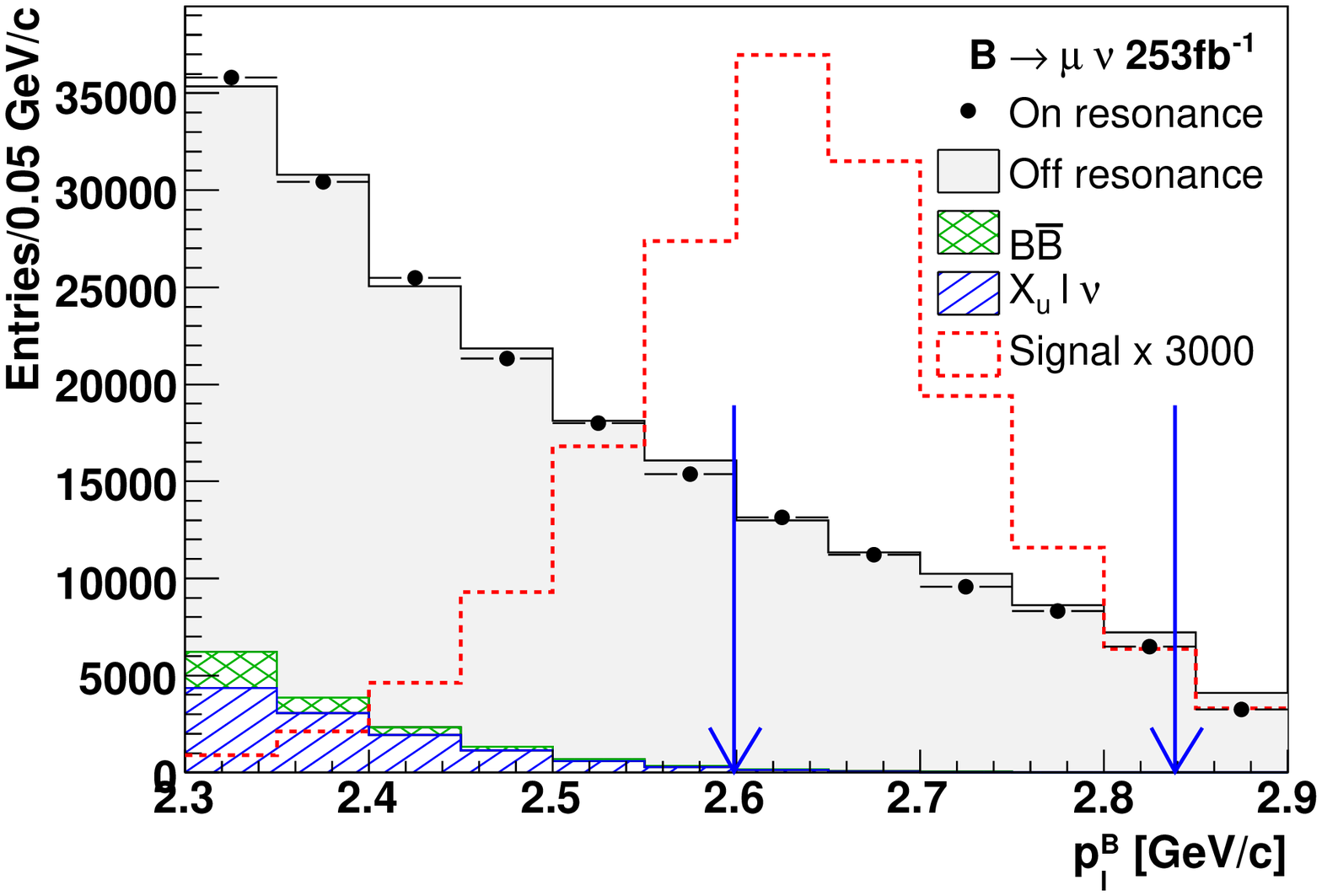}
   \includegraphics[scale=0.34,angle=0]{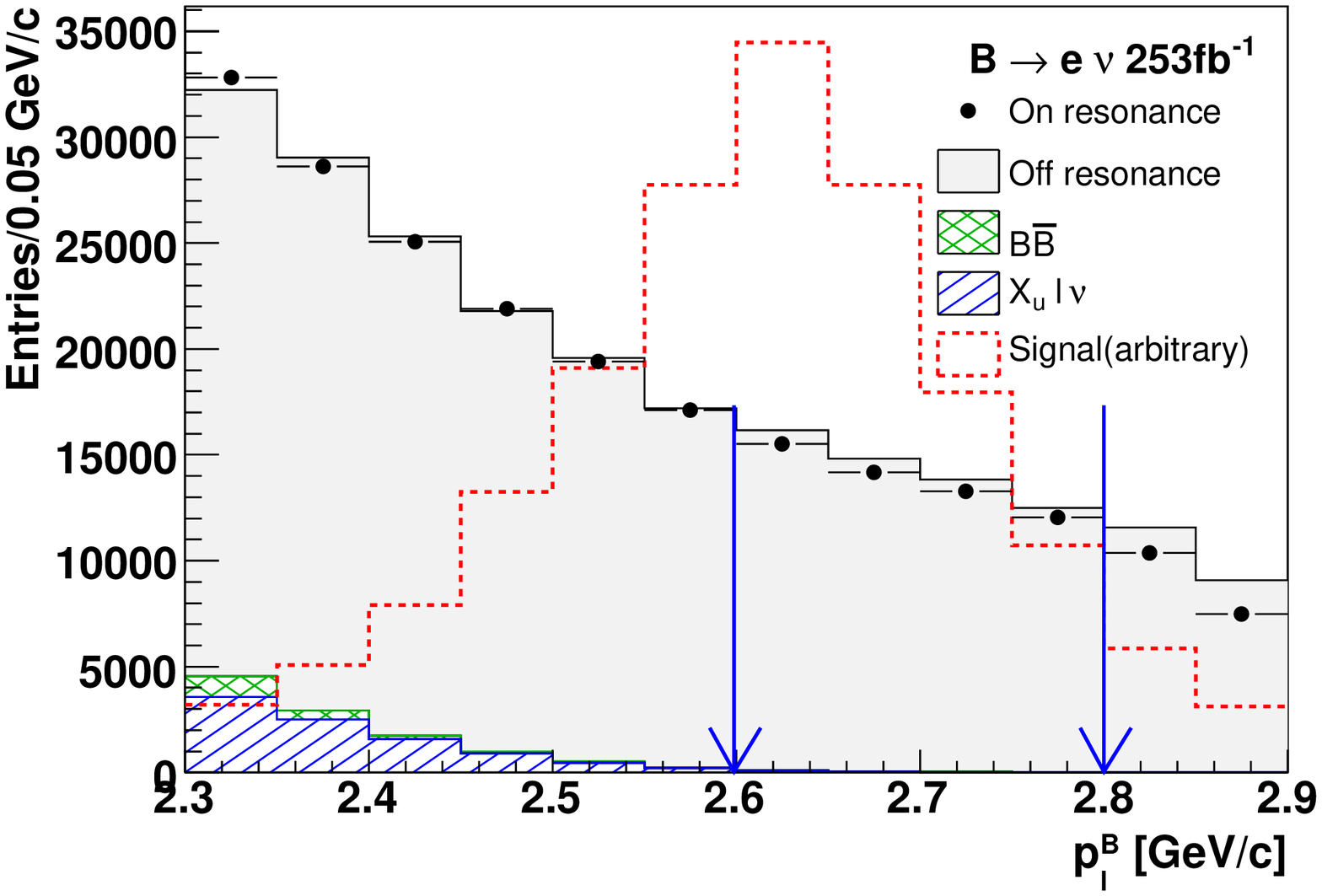}
   \caption{Lepton momentum distributions in the rest frame of
   the signal $B$ meson for the muon (left) and
   electron (right) mode after 
   reconstruction of the companion $B$ meson.
   Points show the on-resonance data, and 
   solid histograms show the expected background due to
   rare $B\to X_{u}\ell\nu$ decays (hatched, from MC); other 
   $B\bar B$ events, principally $B\to X_c \ell \nu$ decays (cross-hatched,
   also from MC); and continuum events (light shaded, taken from scaled
   off-resonance data).
   The dashed histograms represent the signal
   as predicted by the MC with arbitrary normalization.
   }
   \label{plbcm_sig_bg}
  \end{center}
\end{figure}

Figure~\ref{plbcm_sig_bg} shows the measured $p_{\ell}^B$ distributions 
after reconstruction of the companion $B$ meson.
We require $2.6\,\GeV/c < p_\mu^B < 2.84\,\GeV/c$ and
$2.6\,\GeV/c < p_e^B < 2.8\,\GeV/c$ for the signal candidates. 
This requirement removes most of the $B\to X_c \ell \nu$ background.

\begin{figure}[tb]
  \begin{center}
   \includegraphics[scale=0.34,angle=0]{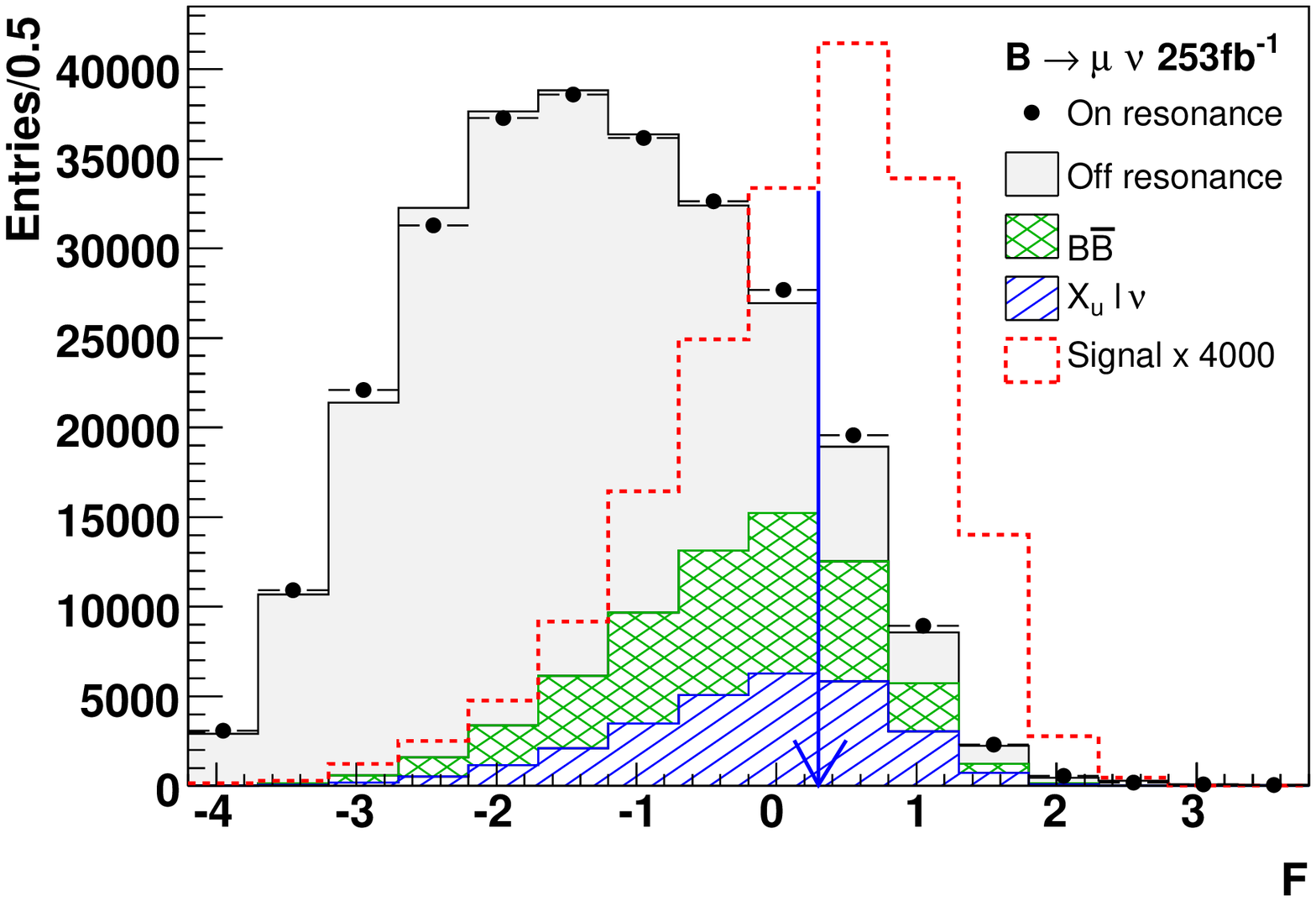}
   \includegraphics[scale=0.34,angle=0]{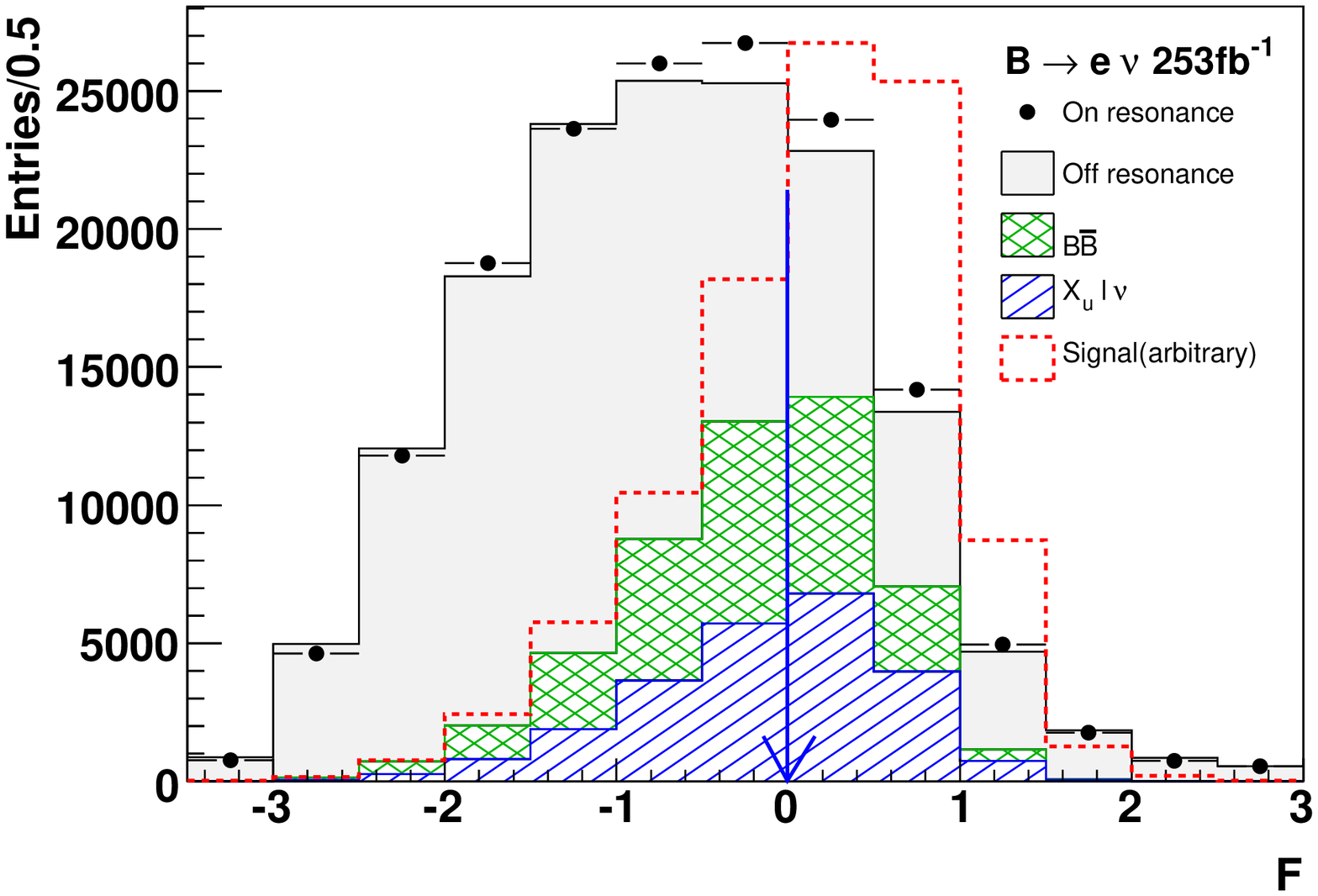}
   \caption{Fisher discriminant distributions for the muon (left) and
   electron (right) mode after $\cos\theta_\ell$ requirements have been applied.
   Points show the on-resonance data, and 
   solid histograms show the expected background due to
   rare $B\to X_{u}\ell\nu$ decays (hatched, from MC); other 
   $B\bar B$ events, principally $B\to X_c \ell \nu$ decays (cross-hatched,
   also from MC); and continuum events (light shaded, taken from scaled
   off-resonance data).
   The dashed histograms represent the signal
   as predicted by the MC with arbitrary normalization.
   }
   \label{fisher}
  \end{center}
\end{figure}

In order to further suppress the continuum background we exploit the
event shape difference
between continuum events, which are jet-like, and $B\bar{B}$ events,
which tend to have a more spherical topology.
Modified Fox-Wolfram moments~\cite{Fox:1978vu,sfw}
are combined into a Fisher discriminant ($F$).
Figure~\ref{fisher} shows the Fisher discriminant distributions
for events that passed the $\cos\theta_\ell$ selection.
We require $F > 0.3$
for the muon mode and $F > 0$ for the electron mode, which retain
approximately $51\,\%$ $(60\,\%)$ of the signal in the signal region 
and remove approximately $99\,\%$ $(95\,\%)$ of the continuum background
in the signal region for the muon (electron) mode.

We determine all selection criteria 
by maximizing $N_S/\sqrt{N_S+N_B}$, where $N_S$ is
the number of signal events expected in the signal region computed
assuming the SM branching fractions and $N_B$ is the number of expected
background events in the signal region from MC.

\begin{figure}[tbp]
  \begin{center}
   \includegraphics[scale=0.34,angle=0]{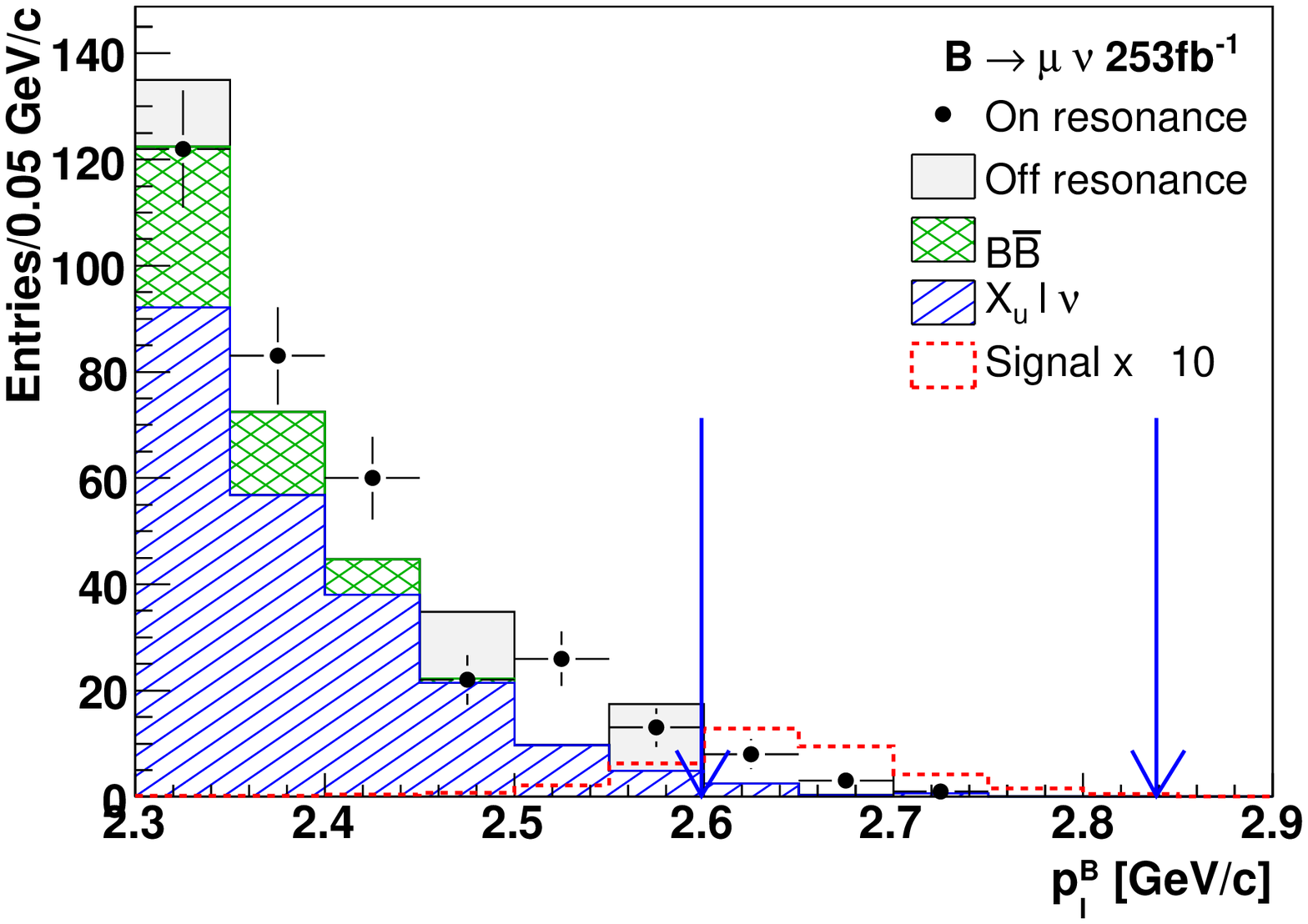}
   \includegraphics[scale=0.34,angle=0]{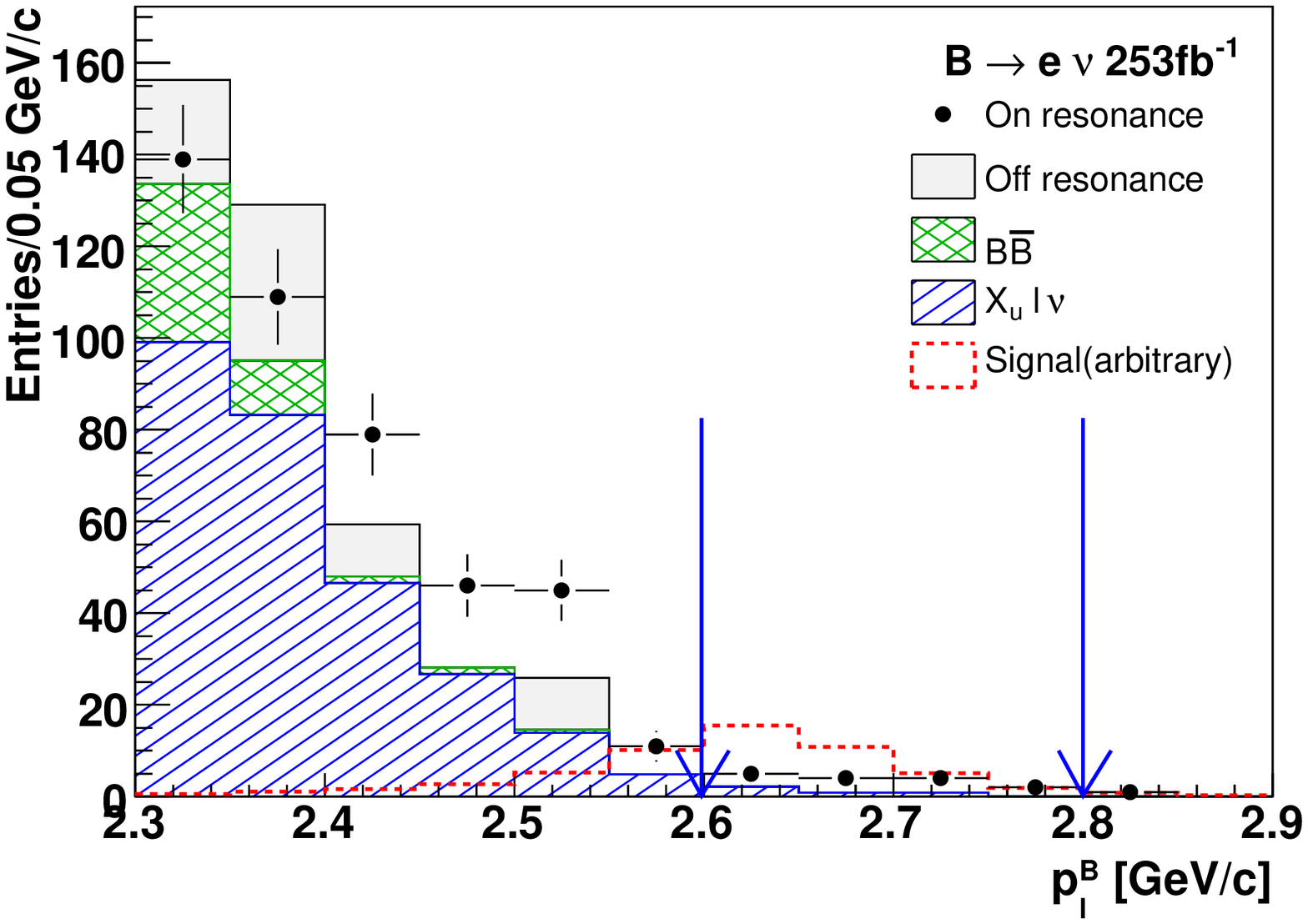}
   \caption{$ p_\ell^B$ distributions for the signal candidates.
   Points show the on-resonance data, and 
   solid histograms show the expected background due to
   rare $B\to X_{u}\ell\nu$ decays (hatched, from MC); other 
   $B\bar B$ events, principally $B\to X_c \ell \nu$ decays (cross-hatched,
   also from MC); and continuum events (light shaded, taken from scaled
   off-resonance data).
   Dashed histograms are MC $B \to \ell \nu$ signals 
   that are obtained by multiplying the SM expectations by a factor of 
   10 for the muon mode and
   by $5\times 10^{6}$ for the electron mode.
   The arrows show the signal regions.
   }
   \label{plbcm_result}
  \end{center}
\end{figure}

After all selection criteria have been applied, the signal selection
efficiency in the signal region is estimated to be 
$\epsilon_{\mu} = 2.18\pm0.06\,\%$ for the muon mode and 
$\epsilon_{e} =2.39\pm0.06\,\%$ for the electron mode.
The efficiencies in the fit region are $3.15\pm 0.07\,\%$ for the 
muon mode and $3.86\pm 0.08\,\%$ for the electron mode.
Figure~\ref{plbcm_result} shows the $ p_\ell^B$ distributions after
all other selections have been applied.
The background that remains in the signal region consists of 
approximately $76\,\%$ continuum and $24\,\%$ $B \to X_u \ell \nu$
according to the off-resonance data and MC studies.
The MC study also indicates that 
the latter exhibits a peaking $M_{\rm bc}$ distribution
with significantly larger width than the signal distribution.

\section{Signal Extraction}
\label{sec:SigExtraction}

The signal yields are extracted from unbinned maximum likelihood fits to
$M_{\rm bc}$ distributions in the fit region.
The signal $M_{\rm bc}$ distribution is parameterized by a Crystal Ball 
function~\cite{Bloom:1983pc} modeled on the signal MC.
The background is described by ARGUS functions~\cite{ARGUS}, with
shape parameters determined from the off-resonance data for the
continuum background and from MC for the $B\bar B$ background.
We do not include the peaking component from $B\to X_u\ell\nu$ in this
fit since an examination of $\Delta E$ sideband data indicates that the
peaking contribution is negligible.
The expected number of background events in the signal region is
estimated by fitting the $M_{\rm bc}$ distribution outside the signal
region to 
a background shape determined from the $B\bar B$ MC and off-resonance data. 
The expected number of background events is $7.4\pm 1.0$ $(13.4\pm 1.4)$
for the muon (electron) mode. Use of the combination of the $B\bar B$ MC
samples and off-resonance data instead of the on-resonance data gives
similar expected number of background events.
The $M_{\rm bc}$ distributions are used as probability density
functions (PDF)
to compute an extended likelihood function defined as follows:
\begin{equation}
 {\mathcal L}(n_s, n_b) =\frac{e^{-(n_s+n_b)}}{N!}\prod^N_{i=1}(n_s f_s(i)+n_b
  f_b(i))
  \label{eq:likelihood}
\end{equation}
where $n_s$ and $n_b$ represent the numbers of signal and background
events in the fit region to be determined in the fit, 
$N$ is the number of observed events, $f_s$ and $f_b$ are the
signal and background PDFs, respectively. 
The negative log likelihood function is minimized using
MINUIT~\cite{James:1975dr} with two free parameters $n_b$ and $n_s$ 
where 
$n_s=\epsilon_{\ell} \times N_{B\bar B} \times 
{\mathcal B}(B^{+}\to\ell^{+}\nu)$ with $\epsilon_{\ell}$ being the
efficiency in the fit region, and $N_{B\bar B}$ the total number of
$B\bar B$ events analysed. We assume the number of the charged and
neutral $B\bar B$ pairs to be equal.

\begin{figure}[tbp]
  \begin{center}
   \includegraphics[scale=0.34,angle=0]{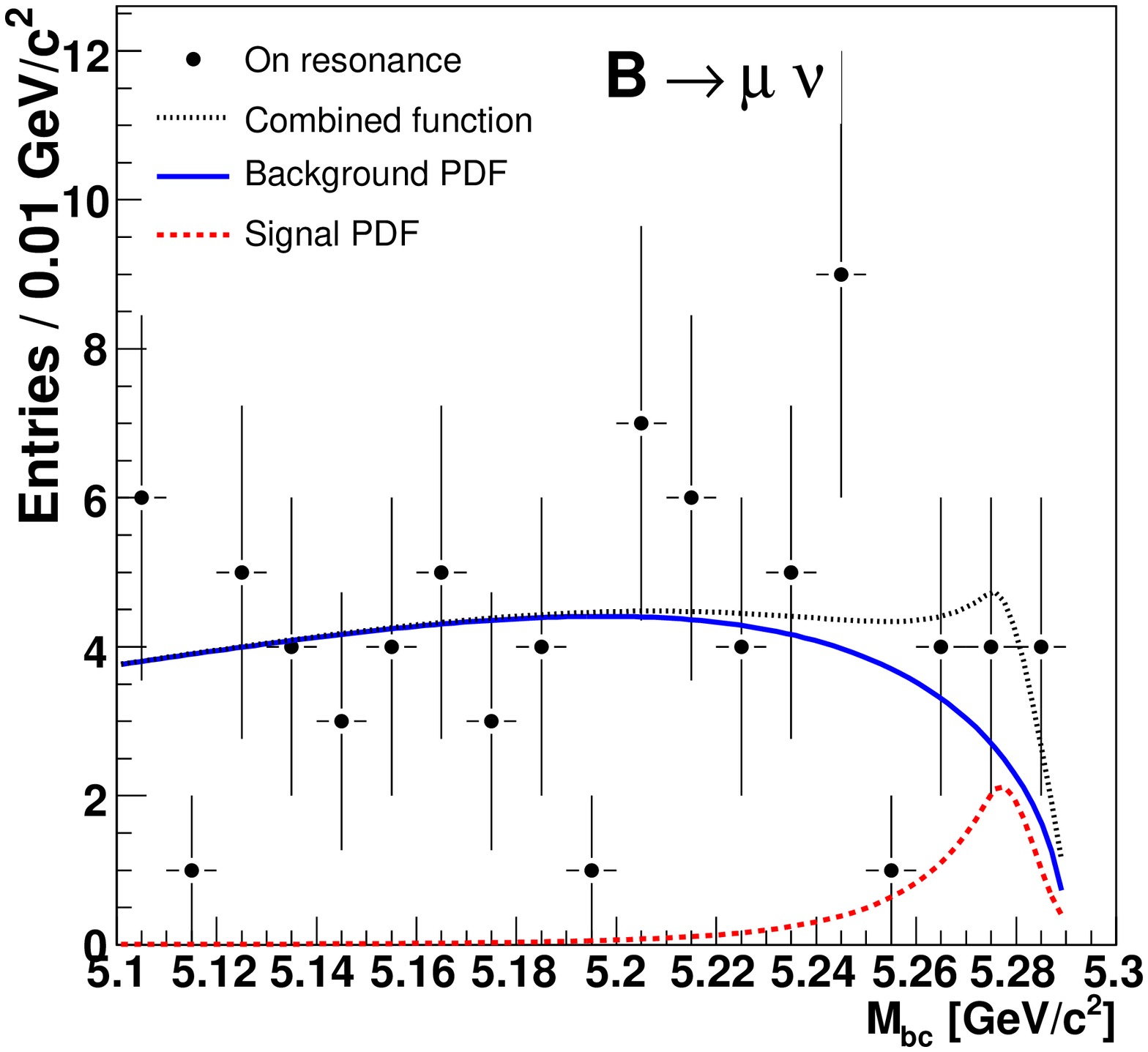}
   \includegraphics[scale=0.34,angle=0]{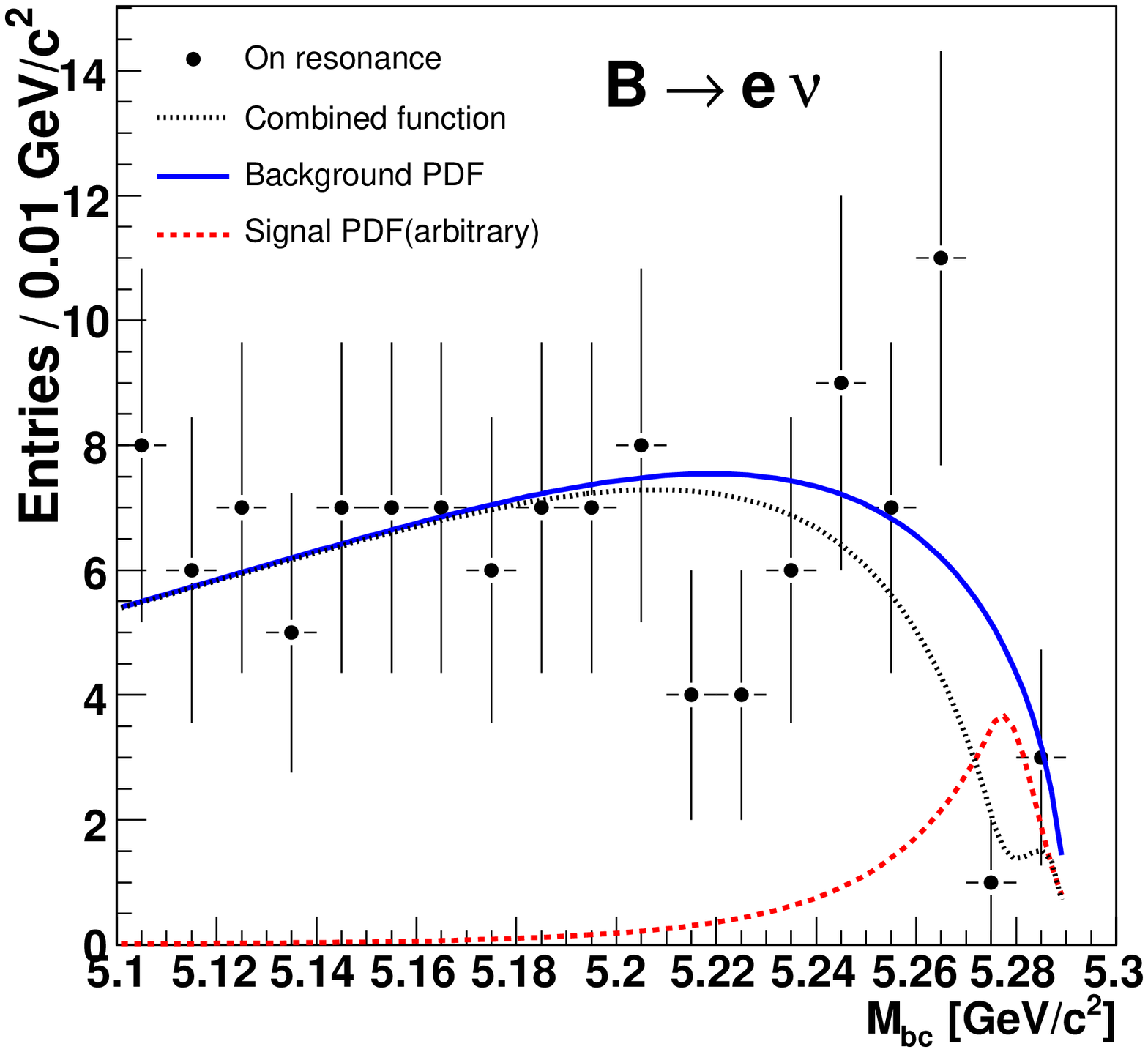}
   \caption{$M_{\rm bc}$ distributions for the events in the $\Delta E$
   signal region, together with the fit results (dotted lines).
   The solid curves are the background contributions. The dashed curves
   are the signal contributions.
   The signal contribution in the electron mode is multiplied
   by a factor of $-4$ to make it visible on the plot.
   }
   \label{bestfit_result}
  \end{center}
\end{figure}

Figure~\ref{bestfit_result} shows 
the $M_{\rm bc}$ distributions of events in the $\Delta E$ signal
region together with the fit results.
We observe 12 (15) events for the muon (electron) mode 
in the signal region.
The signal yield extracted from the fit is $4.1 \pm 3.1$ events for the
muon mode and $-1.8\pm 3.3$ events for the electron mode in the signal
region. For the SM branching fractions, we expect
$2.8\pm0.2$ and $(7.3\pm1.4)\times 10^{-5}$
events for the muon mode and the electron mode, respectively. 
The significance of the signal in the muon mode is $1.3$, which
is defined as $\sqrt{2\ln({\mathcal L}_{\rm max}/{{\mathcal L}_0})}$
where ${\mathcal L}_{\rm max}$ is the likelihood value for the best-fit signal
yield and ${\mathcal L}_0$ is the likelihood value for no signal event.
No excess of events is observed in the electron mode. Selection
efficiencies, expected numbers of events for signal and background and
fit results are summarized in Table~\ref{tab_variable}.

\begin{table}[tbp]
 \begin{center}
  \begin{tabular}{lcc} \hline \hline
   & Muon Mode & Electron Mode \\ \hline
   Signal Efficiency (fit region) &  $3.15\pm 0.07\,\%$ & $3.86\pm 0.08\,\%$ \\
   Signal Efficiency (signal region) &  $2.18\pm0.06\,\%$ & $2.39\pm0.06\,\%$ \\
   Observed in Signal region [events]& 12 & 15 \\
   Expected background [events] & $7.4\pm 1.0$ & $13.4\pm 1.4$ \\
   Signal yield [events] & $4.1\pm3.1$ & $-1.8\pm 3.3$ \\
   Significance   & $1.3$ & -- \\
   SM Prediction [events] & $2.8\pm 0.2$ & $(7.3\pm1.4)\times 10^{-5}$ \\
   \hline \hline
  \end{tabular}  
 \caption{Selection efficiencies, expected numbers of events for signal
 and background and fit results.}
 \label{tab_variable}
 \end{center}
\end{table}


\section{Systematic Uncertainties}
\label{sec:SysUnc}

The main sources of systematic uncertainty in calculating 
${\mathcal B}(B^{+}\rightarrow\ell^{+}{\nu})$ arise from the uncertainties
in the number of $B^{+}B^{-}$ events
($N_{B\bar B} =(276.6\pm3.1)\times 10^{6}$), determination of the
signal efficiency, and parameterization of the $M_{\rm bc}$
distribution for the signal and background.
The uncertainty due to the number of $B\bar B$ pairs is $1.1\%$. 
The uncertainties from the signal efficiencies are: $1\%$ due to
the uncertainties in the track-finding efficiency for the signal,
$4.4\%$ due to the uncertainty in the muon ID efficiency and $1.1\%$ due to
the electron ID efficiency,
$2.3\,\%$ $(2.1\,\%)$ for the muon (electron) mode
from the MC statistics.
We calculate an efficiency correction factor, to account for
differences between MC and data, by analyzing a
control data sample of fully reconstructed $B^{+}\to \bar D^{(*)0}\pi^+$
decays, where we treat the pion as a signal lepton and the $\bar D^{(*)0}$
as the accompanying neutrino.
The event topology is similar to that of the signal events as both are
two-body decays.
The companion $B$ is reconstructed in the control data samples as 
in the signal sample. 
We compare the efficiencies of the control data sample 
and a corresponding MC sample and determine correction factors for the signal efficiencies. 
The correction factors obtained are $1.01\pm0.04$ for the muon mode and 
$1.13\pm0.04$ for the electron mode. 
We apply a correction only to the electron mode,
while uncertainties on the correction factors contribute to the
systematic uncertainties of both modes at the level of $3.6\,\%$.

The uncertainties related to the signal $M_{\rm bc}$ shapes are estimated
by repeating the fits while varying the Crystal Ball function parameters
by their uncertainties; their contribution is 
$6.5\,\%$ for the muon mode and $3.2\,\%$ for the electron mode.
Uncertainties due to the background $M_{\rm bc}$ shapes are estimated
in a similar manner by varying the ARGUS function parameters and yield
$8.1\,\%$ for the muon mode and $15.7\,\%$ for the electron mode.
Table \ref{tab_sysunc} summarizes the contributions to the systematic
uncertainties.
\begin{table}[tbp]
 \begin{center}
 \begin{tabular}{llcc} \hline \hline
Sources   &           &   Muon Mode     & Electron Mode      \\
\hline
$N_{B\bar B}$  &    & 1.1\%   & 1.1\% \\
Signal Efficiency      & Lepton ID        & 4.4\%   & 1.1\% \\
                       & Tracking         & 1.0\%   & 1.0\% \\
                       & MC statistics    & 2.3\%   & 2.1\% \\
                       &$B^+\to D^0\pi^+$ & 3.6\%   & 3.6\% \\
  $M_{\rm bc}$ Shape & Signal             & 6.5\%   & 3.2\% \\ 
                     & Background       & 8.1\%   & 15.7\% \\
\hline
  Total &  & 12.2\% & 16.7\% \\ 
\hline \hline
 \end{tabular}
 \caption{Summary of systematic uncertainties}
 \label{tab_sysunc}
 \end{center}
\end{table}

\section{Limits on Branching Fractions}
\label{sec:LimitsOnBr}
\begin{figure}[hbtp]
  \begin{center}
   \includegraphics[scale=0.34,angle=0]{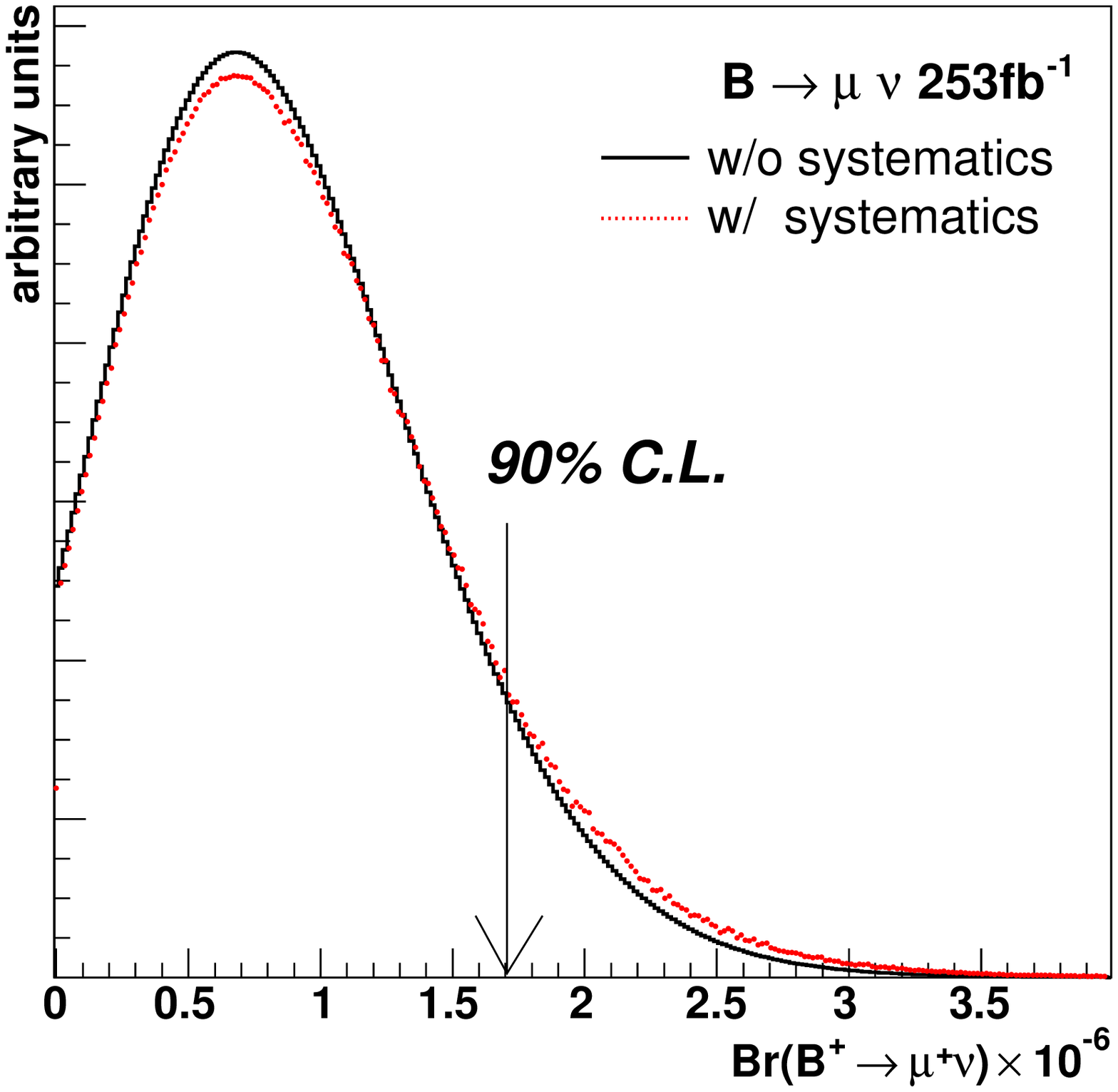}
   \includegraphics[scale=0.34,angle=0]{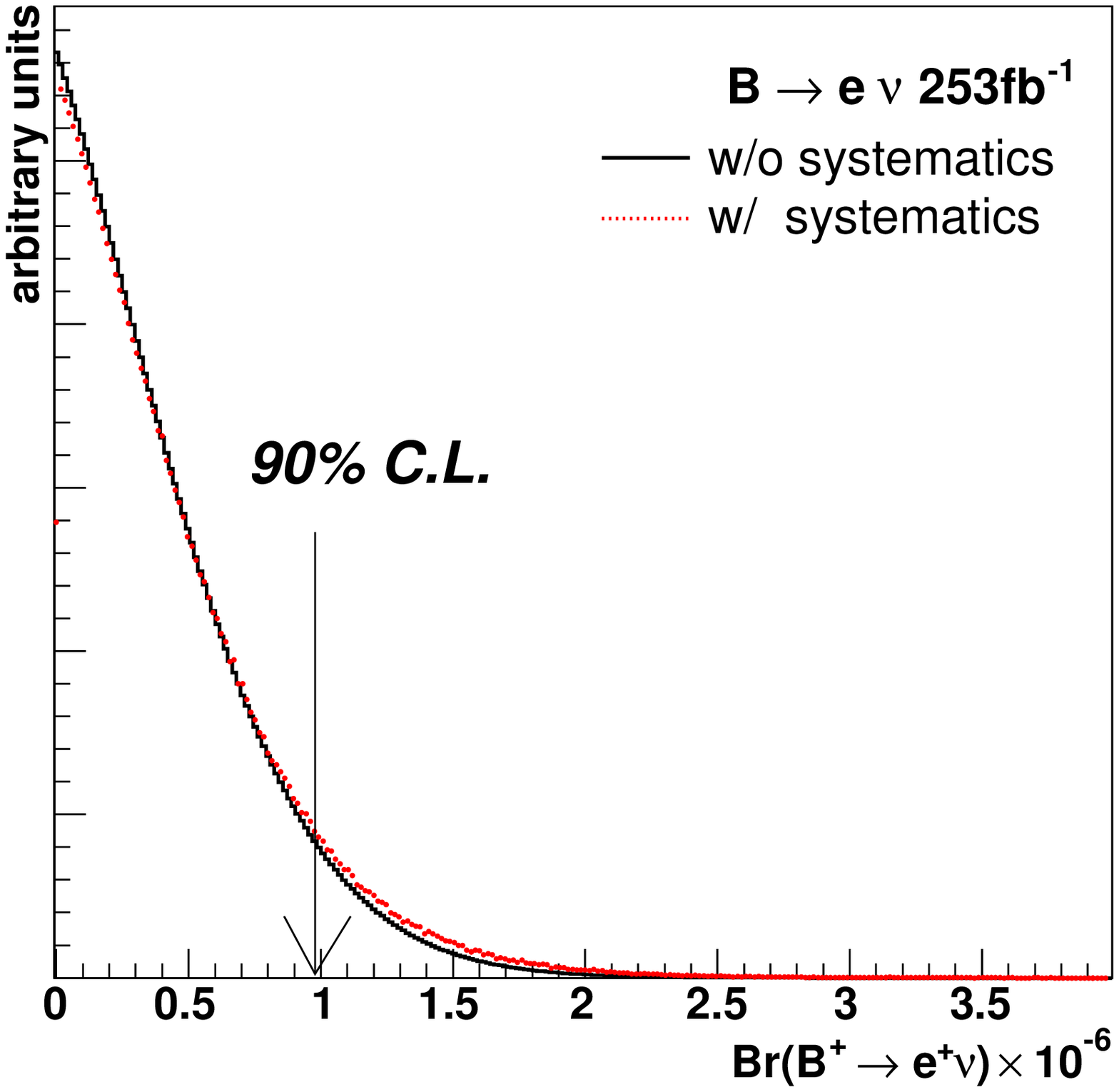}
   \caption{Likelihood function dependence on the branching fractions.
   The solid and dotted curves represent the likelihood functions
   without and with inclusion of systematic uncertainties, respectively.
   The arrows indicate the upper limits on the 
   branching fractions at  $90\%$ confidence level.}
   \label{likelihood_dist}
  \end{center}
\end{figure}

Figure~\ref{likelihood_dist} shows the dependence of the likelihood functions
on the branching fractions, ${\mathcal L (B)}$, before and after inclusion of 
the systematic uncertainties. 
To account for systematic uncertainties in the calculation of the upper
limits for each mode, we convolve the likelihood function 
${\mathcal L (B)}$ with a Gaussian distribution, where the sigma of the
Gaussian corresponds to the size of the systematic uncertainty in this mode.
The $90\,\%$ confidence level for
the upper limit on the branching fraction, ${\mathcal B}_{90}$, is defined by 
$0.9 = {\int^{{\mathcal B}_{90}}_0 {\mathcal L}({\mathcal B})d{\mathcal
B}}/{\int^{\infty}_0 {\mathcal L}({\mathcal B})d{\mathcal B}}$.

We obtain the following upper limits on the
branching fractions at the $90\,\%$ confidence level:
\begin{eqnarray}
 {\mathcal B}(B^+\to\mu^+\nu_\mu) &<& 1.7 \times 10^{-6} ~~(90\,\%~\mathrm{C.L.}) \\
 {\mathcal B}(B^+\to e^+ \nu_e) &<& 9.8 \times 10^{-7} ~~(90\,\%~\mathrm{C.L.}),
\end{eqnarray}
including the effect of the systematic uncertainties. 
The expected sensitivities on the upper limits with the present dataset,
computed using toy MC studies with a null signal hypothesis, are 
$1.0\times 10^{-6}$ for the muon mode and
$1.1\times 10^{-6}$ for the electron mode.

\section{Conclusion}
\label{sec:conlusion}

We have searched for the purely leptonic decays
$B^{+}\to\mu^{+}\nu_\mu$ and $B^{+}\to e^{+}\nu_{e}$
using data collected by the Belle detector at the KEKB $e^{+}e^{-}$ asymmetric-energy
collider.
We have found no evidence of signal in either decay mode.
We set upper limits on the branching fractions:
${\mathcal B}(B^+\to\mu^+\nu_\mu) < 1.7 \times 10^{-6}$
and ${\mathcal B}(B^+\to e^+ \nu_e) < 9.8 \times 10^{-7}$ at $90\,\%$
confidence level. These limits are the most stringent to date and
improve the previous published
limits~\cite{babar-2004,CLEO-1995} by a factor of 4 
for $B^{+}\to\mu^+\nu_\mu$ and 15 for $B^+\to e^+\nu_e$.

\vspace{1.0cm}
We thank the KEKB group for the excellent operation of the
accelerator, the KEK cryogenics group for the efficient
operation of the solenoid, and the KEK computer group and
the National Institute of Informatics for valuable computing
and Super-SINET network support. We acknowledge support from
the Ministry of Education, Culture, Sports, Science, and
Technology of Japan and the Japan Society for the Promotion
of Science; the Australian Research Council and the
Australian Department of Education, Science and Training;
the National Science Foundation of China and the Knowledge
Innovation Program of the Chinese Academy of Sciences under
contract No.~10575109 and IHEP-U-503; the Department of
Science and Technology of India; 
the BK21 program of the Ministry of Education of Korea, 
the CHEP SRC program and Basic Research program 
(grant No.~R01-2005-000-10089-0) of the Korea Science and
Engineering Foundation, and the Pure Basic Research Group 
program of the Korea Research Foundation; 
the Polish State Committee for Scientific Research; 
the Ministry of Science and Technology of the Russian
Federation; the Slovenian Research Agency;  the Swiss
National Science Foundation; the National Science Council
and the Ministry of Education of Taiwan; and the U.S.\
Department of Energy.


\end{document}